\documentclass[]{elsarticle}
\usepackage{float}
\usepackage{amssymb}
\usepackage{amsmath}
\usepackage{color,soul}
\usepackage{lmodern}
\usepackage[rgb]{xcolor}
\usepackage[author={Pablo S. Cornaglia}]{pdfcomment}
	\usepackage{multirow}
\usepackage[utf8]{inputenc}
\journal{JMMM}

\usepackage{array} 
\usepackage[T1]{fontenc}
\newcolumntype{C}[1]{>{\centering\arraybackslash}p{#1}}

\graphicspath{{./figs/}}

\begin{document}
\begin{frontmatter}
\title{Lattice specific heat for the RMIn$_5$ (R = Gd, La, Y; M = Co, Rh) compounds: non-magnetic contribution subtraction}
\author[CAB,CONICET]{Jorge I. Facio}
\author[CAB,CONICET]{D. Betancourth}
\author[CAB,CONICET]{N. R. Cejas Bolecek}
\author[SARMIENTO]{G. A. Jorge}
\author[CAB,CONICET]{Pablo Pedrazzini}
\author[CAB,CONICET]{V. F. Correa}
\author[CAB,CONICET]{Pablo S. Cornaglia}
\author[CAC,CONICET]{V. Vildosola}
\author[CAB,CONICET]{D. J. Garc\'{\i}a}
\address[CAB]{Centro At{\'o}mico Bariloche and Instituto Balseiro, CNEA, 8400 Bariloche, Argentina}
\address[CONICET]{Consejo Nacional de Investigaciones Cient\'{\i}ficas y T\'ecnicas (CONICET), Argentina}
\address[SARMIENTO]{Instituto de Ciencias, Universidad Nacional de General Sarmiento, Buenos Aires, Argentina}
\address[CAC]{Centro At{\'o}mico Constituyentes, CNEA, 1650 San Mart\'{\i}n, Buenos Aires, Argentina}

\begin{abstract}
We analyze theoretically a common experimental process 
 used to obtain the magnetic contribution to the specific heat of a given magnetic material. In the procedure, the specific heat of a non-magnetic analog is measured and used to subtract the non-magnetic contributions, which are generally dominated by the lattice degrees of freedom in a wide range of temperatures. 
We calculate the lattice contribution to the specific heat for the magnetic compounds GdMIn$_5$ (M = Co, Rh) and for the non-magnetic YMIn$_5$ and LaMIn$_5$ (M = Co, Rh),  using density functional theory based methods.
We find that the best non-magnetic analog for the subtraction depends on the magnetic material and on the range of temperatures.
While the phonon specific heat contribution of YRhIn$_5$ is an excellent approximation to the one of GdCoIn$_5$ in the full temperature range, for GdRhIn$_5$ we find a better agreement with LaCoIn$_5$, in both cases, as a result of an optimum compensation effect between masses and volumes. 
We present measurements of the specific heat of the compounds GdMIn$_5$ (M = Co, Rh) up to room temperature where it surpasses the value expected from the Dulong-Petit law. We obtain a good agreement between theory and experiment when we include anharmonic effects in the calculations. 
\end{abstract}

\begin{keyword}
\PACS{75.50.Ee, 63.20.D-, 71.20.-b, 65.40.De}
\end{keyword}

\end{frontmatter}
\section{Introduction}
Specific heat measurements are a mighty tool to analyze and unveil quantum critical points, superconducting or magnetic phase transitions and heavy fermion behavior \cite{Sereni19911}.
The temperature behavior of the specific heat can be used to extract microscopic parameters like the effective mass of the electric carriers and information about the order of a phase transition.  Different degrees of freedom typically contribute to the specific heat, and when a set of degrees of freedom is weakly coupled to the others, it becomes possible to analyze its contribution in a separate way. 
In magnetic transitions, extracting the magnetic specific heat allows to obtain information about the magnetic moments involved, the nature of the transition, and can be used to test theoretical models and methods.
For some materials and temperature ranges, one contribution may dominate over the others or have a characteristic temperature dependence allowing to extract useful information about it directly from the experiment, e.g. the electronic contribution at low temperatures. Other contributions, however, may be difficult to disentangle because they have a temperature dependence similar to other contribution or because they are small. This is the case of the magnetic contribution which, in some compounds, is small and has a low temperature behavior with a temperature dependence of the same form as the phonon one.

A common experimental procedure to obtain the magnetic contribution to the specific heat is to subtract the specific heat of an 
analogous non-magnetic compound \cite{mori1985new,lora2006magnetic,lora2009doping,VanHieu2007,duque2011field,matsumoto2011local,singh2012relevance,adriano2014physical,vcermak2014magnetic}. Since the non-magnetic contribution is dominated in general by the lattice degrees of freedom, it is not obvious a priory why this procedure should work. 
While the analogous compound has the same crystal structure as the magnetic one, the mass and size of the atomic constituents differ leading to a change in the frequency of the phonon modes and, as a consequence, to a different temperature behavior of the lattice specific heat. 

In the present work, we analyze theoretically the above mentioned experimental subtraction procedure. To that aim, we study the lattice dynamics and its contribution 
to the specific heat of the magnetic compounds GdMIn$_5$ (M = Co, Rh) and analyse differences and similarities with the non-magnetic 
analogous materials  RMIn$_5$ (R = Y, La; M = Co, Rh). 
These materials belong to the so-called 115 family of compounds where R can be a 4f or 5f atomic element. 
The multi-orbital character of the 115 materials, with important crystal-field and spin-orbit (SO) effects 
and, in some cases, a strong hybridization between the f states with the conduction bands, produce fascinanting  
physical properties like unconventional superconductivity, heavy fermion behavior, and different magnetic phases that are very complex to assess.

The Gd-based compounds  simplify the physics of this family because the Gd$^{+3}$ ion has zero orbital angular momentum, small SO effect
and negligible 4f hybridization with the conduction electrons. This makes them a good starting point to study the low temperature 
electronic and magnetic properties of the 115 series, and get insights about its more complex compounds. 
Interestingly, GdCoIn$_5$ has a more two-dimensional magnetic behavior than its Rh and Ir counterparts which leads to a lower N\'eel temperature~\cite{facio2015co}. This may explain the lower N\'eel temperatures observed for the Co based compounds across the 4f series, and help understand the higher superconducting critical temperature in the Co based Ce and Pu 115 compounds. 

In order to properly describe the specific heat of these compounds, at least  three contributions must be considered~\cite{Betancourth2015,facio2015co}: 
i) the lattice dynamics, ii) the electronic bandstructure,  and iii) the magnetic interactions. 
We perform total-energy {\it ab initio} calculations and obtain the phonon spectra and specific heat of magnetic and non-magnetic compounds of the 115 family.
Our results indicate that the best non-magnetic analog for the subtraction of the phonon contribution to the specific heat will have to satisfy a compensation effect between masses and volumes of its constituents. However this is a delicate balance that will depend on the magnetic material and also  on the temperature range that makes a hard task the formulation of a general rule. As it is shown in this work, Density Functional Theory (DFT) calculations can help to obtain the actual phonon background and/or choose the best non-magnetic compound to subtract this background experimentally.

We also measure the specific heat of GdCoIn$_5$ and GdRhIn$_5$ samples up to room temperature and analyze the calculated phonon contribution for the subtraction. This subtraction has associated difficulties due to limitations of the calculations: these are done at zero temperature and constant volume in the harmonic approximation. The measured specific heats surpass the Dulong-Petit law high temperature value to an extent that cannot be explained by electronic or magnetic contributions in these materials, and that we attribute to anharmonic effects in the lattice.
We find that including a semiphenomenological correction to the calculated phonon specific heat, using a single adjustable parameter, we obtain a good description of the anharmonic effects at high temperature, allowing the phonon subtraction.

The rest of the paper is organized as follows: Sec. \ref{sec:lattice} describes the {\it ab initio} calculations of the crystal structure, the phonon spectrum and the lattice contribution to the specific heat. Section \ref{sec:exp} details the sample preparation and the specific heat measurements. In Sec. \ref{sec:comp} we compare the experimental results with the theory and in Sec. \ref{sec:concl} we present our conclusions.

\section{Theoretical results for the lattice properties}\label{sec:lattice}
We perform first-principles calculations based on Density Functional Theory using the full-potential augmented plane waves + local orbitals method as implemented in the \textsc{WIEN2K} code \cite{wien2k}. 
The results that we present here have been obtained within the generalized gradient approximation (GGA) for the exchange and correlation functional~\cite{Perdew1996a}.
In order to better account for the local Coulomb repulsion in the Gd $4f$ shell, we use GGA+U method. This method has already been used in previous works to study other compounds of the RMIn$_5$ family \cite{Piekarz2005,PhysRevLett.96.237003,Zhu2012}. Due to the localized character of the $4f$ electrons the fully localized limit was used 
for the double counting correction \cite{Anisimov1993}. The local Coulomb and exchange interactions are described with a single effective local repulsion $U_{eff} = U - J = 6 eV$ as in bulk Gd \cite{Yin2006,Petersen2006}. Taking into account that the atomic ground state of Gd$^{3+}$ has L=0 we decided to neglect the spin-orbit coupling in these systems. Both the lattice parameters and the internal positions were fully relaxed considering 1200 $\mathbf{k}$ points in the Brillouin zone and a plane-wave cut-off $K_{max}$ determined by fixing $RK_{max}$ =8.5, being $R$= 2.41 a. u., the muffin-tin radii of the In atoms. We verified that the calculations were converged in the number of 
$\mathbf{k}$ points and plane waves.

\subsection{Crystal Structure}\label{sec:crystal}
\begin{figure}[t]
    \begin{center}
        \includegraphics{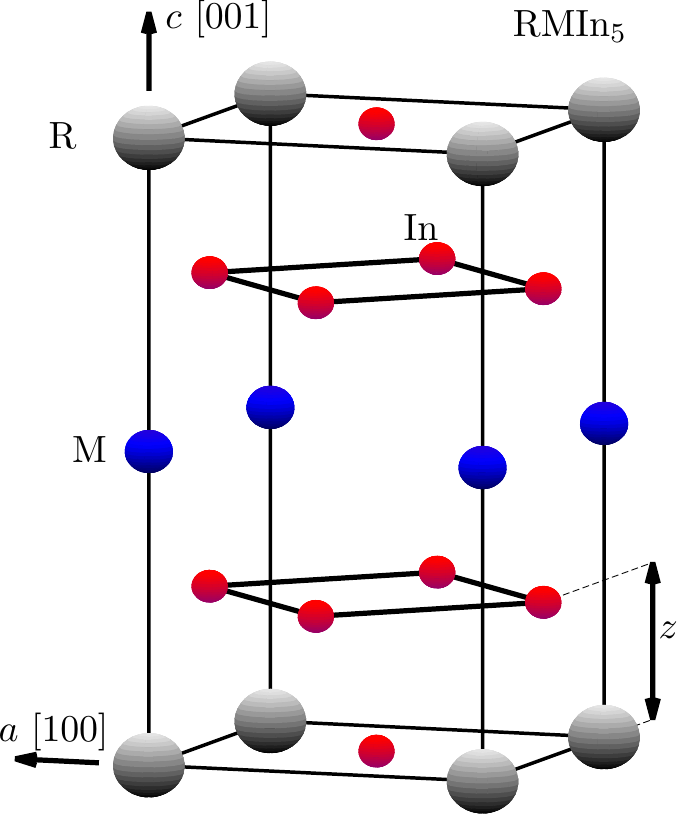}
    \end{center}
    \caption{(Color online) Crystal structure for the 115 compounds. The In atoms are represented by red spheres, the transition metal M by blue spheres and the R element by grey spheres.}
    \label{fig:unitcell}
\end{figure}
The materials under study crystallize in the tetragonal HoCoGa$_5$ structure ($P4/mmm$ space group), which is fully defined by the lattice parameters $a$ and $c$ plus one internal coordinate $z$ that defines the position of one type of the In ions (see Fig. \ref{fig:unitcell}). 
To obtain the equilibrium value of these parameters we iteratively find the values of $a$, $c$ and $z$ that minimize the total energy $E(a,c,z)$.
The threshold energy and forces used to determine convergence are $0.01 mRy$ and  $0.1 mRy/a.u.$, respectively.

\begin{table}[h]
\begin{tabular}{ |l|l|l|l|l| }
\hline
\multirow{4}{*}{}  &  & Experiment (\AA) & GGA (\AA)& \% diff. \\
\hline
\multirow{4}{*}{YCoIn$_5$}
 & $a$   & $4.551^a$    & 4.593  & 0.9 \\ 
 & $c$   & $7.433^a$    & 7.529  & 1.3 \\ 
 & $z/c$ &              & 0.3067  &  \\\hline 
\multirow{4}{*}{YRhIn$_5$}  
 & $a$   & 4.601$^b$    & 4.675  & 1.6 \\ 
 & $c$   & 7.435$^b$    & 7.516  & 1.6 \\ 
 & $z/c$ & 0.3001$^b$   & 0.3011  & 0.1  \\\hline 
\multirow{4}{*}{LaCoIn$_5$}
 & $a$   & 4.639(1)$^c$ & 4.680  & 0.9 \\ 
 & $c$   & 7.615(4)$^c$ & 7.700  & 1.1 \\ 
 & $z/c$ &0.31134(9)$^c$& 0.3120  & 0.3  \\\hline 
\multirow{4}{*}{LaRhIn$_5$}  
 & $a$   & 4.673$^b$    & 4.751  & 1.6 \\ 
 & $c$   & 7.590$^b$    & 7.694  & 1.4 \\ 
 & $z/c$ & 0.3078$^b$   & 0.3069  & 0.3  \\\hline 
\multirow{4}{*}{GdCoIn$_5$}  
 & $a$   & 4.568(1)$^d$ & 4.606  & 0.9 \\ 
 & $c$   & 7.4691(7)$^d$& 7.559  & 1.3 \\ 
 & $z/c$ &              & 0.3077  &  \\\hline
\multirow{4}{*}{GdRhIn$_5$}
 & $a$   & 4.606$^b$    & 4.685  & 1.7 \\ 
 & $c$   & 7.439$^b$    & 7.555  & 1.6 \\ 
 & $z/c$ & 0.3025$^b$   & 0.3024  & 0.05   \\\hline
\multirow{4}{*}{GdIrIn$_5$}
 & $a$   & 4.622(4)$^e$ & 4.700  & 1.7 \\ 
 & $c$   & 7.413(8)$^e$ & 7.545  & 1.8 \\ 
 & $z/c$ &              & 0.3019  &  \\\hline
\end{tabular}
\caption{Structural parameters (in \AA) for the 115 compounds analyzed. The rightmost column indicates the relative differences (in \%) between theory and experiment. Superscipt letters correspond to References: $a=$\cite{Kalychak199980}, $b=$\cite{VanHieu2007}, $c=$\cite{Macaluso2002}, $d=$\cite{Betancourth2015}, and $e=$\cite{Pagliuso2001}.}
\label{struct}
\end{table}

Table~\ref{struct} presents the experimental and calculated values of the lattice constants for the different compounds studied. In general, the structural parameters calculated with GGA are about $1$--$1.5\%$ larger than the experimental observations, with a concomitant overestimation of the volume of around $3$--$5\%$. The calculations, however, capture accurately the relative changes in the lattice constants as the R and M elements are replaced.
An interesting feature observed in the experiments and reproduced by the DFT-based calculations is the reduction of the lattice parameter $c$ and the increase of $a$ as the transition metal is changed from Co to Rh and from Rh to Ir. 
Both the theory and the experimental data show that the structures of the La compounds have larger volumes than the Y and Gd ones. As we show below, this is associated with an overall softening of the optical phonon modes in the La compounds. 

\subsection{Elastic properties}\label{elastic_properties}
We calculate from first-principles the phonon dynamical matrix in the harmonic aproximation using the Parlinksi-Li-Kawasoe method (for further details see Ref. \cite{PhysRevLett.78.4063}) as implemented in the \textsc{phonopy} code \cite{PhysRevB.78.134106}.
Basically, for the crystal structure under study, a set of independent atomic displacements (not related by symmetries) is defined. 
For each of those displacements, a supercell DFT calculation is performed until selfconsistency. 
As a result, Hellmann-Feynmann forces on all atoms within the supercell are obtained. These allow to determine the force constant matrices by singular value decomposition.
In our case there are 9 displacements not related by symmetry (see Ref. \cite{Piekarz2005}) that can be realized 
in a 2x2x1 supercell. To improve the precision, for each of these displacements, we also consider the displacement in the opposite direction. The size of the atomic displacements is around 1$\%$ of the lattice parameter.


\subsection{Phonons}
We calculated the zero-temperature phonon spectrum and phonon density of states (DOS) of the different compounds in order to estimate the lattice contribution to the specific heat. Within the harmonic approximation, the phonon spectrum is temperature independent and we can calculate the contribution to the specific heat as a function of the temperature from the zero-temperature phonon DOS. As we discuss below, at high enough temperatures, anharmonic effects do become relevant and a related contribution to the specific heat needs to be considered for a quantitative comparison with the experimental data. An additional correction is also needed because the experiments are performed at constant pressure, while the calculations are done for the T=0 volume. 

\begin{figure}[!t]
  \centering
    \includegraphics[width=\textwidth]{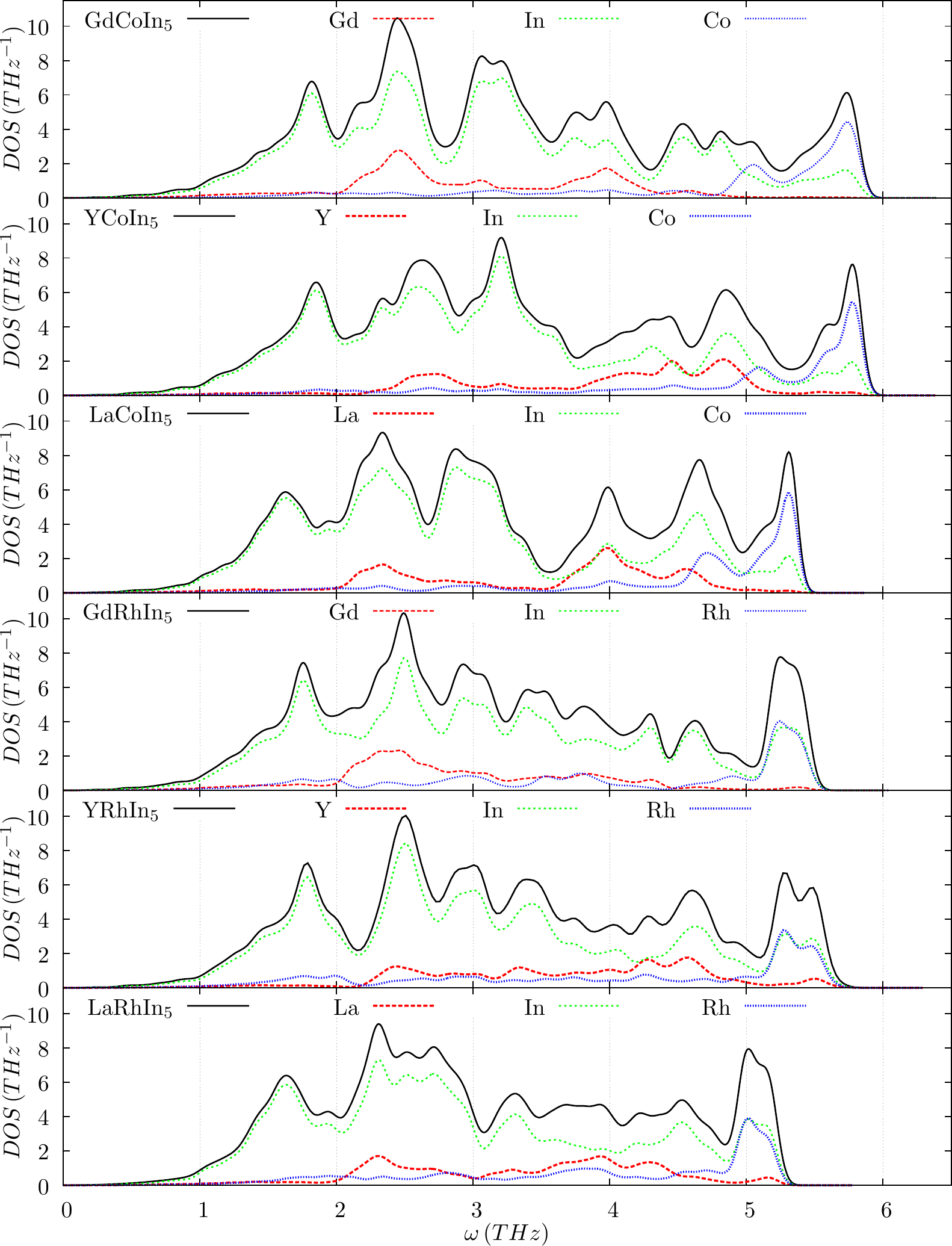}
  \caption{Phonon density of states for the different compounds studied. The total density of states of each compound is presented with a solid black line. The weighted DOS acording to the participation of the different ions in the phonon modes are presented with other line styles and colors as indicated on each panel.
	}
  \label{DOS_phon_Co}
\end{figure}

Figure \ref{DOS_phon_Co} presents the total and partial phonon DOS for the different compounds studied. The partial DOS for a given ion is calculated weighting the spectral density of the phonon modes according to the contribution 
 of the ion. The transition metal atoms participate mainly on the high frequency optical modes, the R atoms take part 
 in modes in the middle of the optical spectrum, and 
all optical modes have a significant In character 
\footnote{ A detailed analysis of the different modes for this family of compounds was presented by Piekarz {\it et al.} in Ref. \cite{Piekarz2005}}. 
When the Gd atom is replaced by Y, the high and low frequency parts of the optical spectrum are only slightly modified, and there is a redistribution of spectral weight in the center of the optical phonon spectrum with a shift of the weight to higher energies due to the hardening of the modes associated with the reduction of the mass of the R ion. 
When Gd is replaced by La, however, there is an overall softening of the optical phonon modes (signaled by a shift of the spectrum to lower frequencies). The converse situation, however, would be naively expected due to the larger mass of the Gd ion. 
The observed total DOS difference stems from a reduction in the lattice elastic constants associated with the Co and In atoms in the La compound. Nevertheless, the partial phonon DOS analysis reveals that the modes in which the R ion is involved are shifted to higher energies.
This weakening of the elastic constants could be related to the increase of the equilibrium volume of the unit cell when Gd is replaced by La (see Table \ref{struct}).
When Co is replaced by Rh, the main effect on the phonon spectrum is a softening of the highest energy optical mode which involves mainly the transition metal atoms. The spectral weight of the lowest energy optical modes, which are mainly associated with the In ions, remains essentially unchanged and there is a redistribution of spectral weight in the middle of the spectrum. A similar behavior is obtained for the La and Y compounds when the transition metal is replaced. 

As we will see below, the best approximation to the phonon contribution to the specific heat of GdCoIn$_5$ is obtained with YRhIn$_5$. Replacing Co by Rh in the YCoIn$_5$ compound, there is a softening of the modes associated with the transition metal due to the higher mass of the ion. This softening partially compensates the hardening associated with the replacement of Gd by Y in the GdCoIn$_5$ compound.

The case of GdRhIn$_5$ is quite different, the best candidate to reproduce its phonon contribution to the specific heat, among the studied compounds, is LaCoIn$_5$. The hardening of the modes expected when Gd is replaced by La and Rh by Co is partially compensated by the weakening of the elastic constants in the La compound.
\subsection{Lattice contribution to the Specific heat}
We now analyze the phonon contribution to the specific heat for GdCoIn$_5$ and GdRhIn$_5$ to compare with the nonmagnetic compounds 
YRhIn$_5$, YCoIn$_5$, LaRhIn$_5$, and LaCoIn$_5$. 

\begin{figure}[t]
  \centering
    \includegraphics[width=\textwidth]{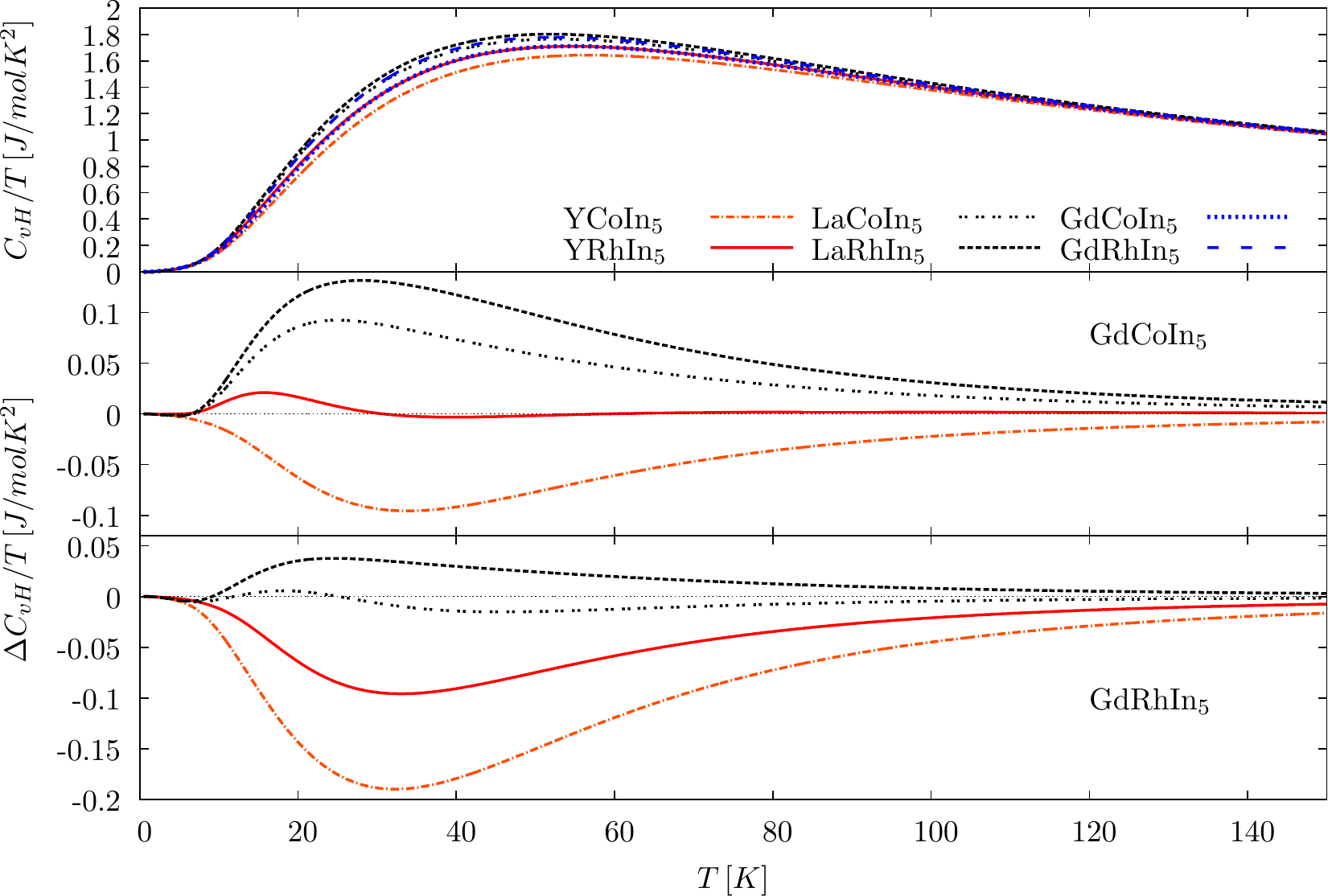}
		\caption{Top panel: Calculated harmonic lattice contribution to specific heat $C_{vH}/T$ for the different 115 compounds studied. Central panel: $C_{vH}/T$ difference between the non-magnetic compounds and GdCoIn$_5$. Central panel: $C_{vH}/T$ difference between the non-magnetic compounds and GdRhIn$_5$.}
  \label{cv_todos}
\end{figure}

The top panel of Fig. \ref{cv_todos} presents $C_{vH}/T$ vs $T$ for all the compounds studied. The largest differences between the specific heats 
 are obtained in the range of temperatures $10K<T<60K$, that includes the N\'eel temperatures of the magnetic compounds. 
Precisely in this range of temperatures the largest contributions to the magnetic specific heat are obtained and is critical to perform an accurate subtraction of the phonon contribution. 
The result of subtracting the $C_{vH}/T$ values for GdCoIn$_5$ to the corresponding for the non-magnetic compounds are presented in central panel.
There is a strong resemblance between the $C_{vH}/T$ vs $T$ curves of GdCoIn$_5$ and YRhIn$_5$. Interestingly, the unit cell masses and volumes of these two compounds differ only by $3\%$ and $0.77\%$, respectively.
While the difference in mass is slightly smaller between GdCoIn$_5$ and LaRhIn$_5$ (2.3\%) these compounds have a much larger difference in volume (5\%). The LaCoIn$_5$ optical modes are softer and its $C_{vH}(T)$ curve 
lies above the GdCoIn$_5$ one at intermediate temperatures ($20K<T<200K$). The opposite situation occurs for YCoIn$_5$ which has a unit cell volume similar to GdCoIn$_5$ but has a unit cell mass significantly lower.
 For GdCoIn$_5$ the best choice found for the experimental subtraction is YRhIn$_5$ while for GdRhIn$_5$ our calculations indicate that LaCoIn$_5$ should be used (see bottom panel of Fig. \ref{cv_todos}).
The entropy difference ($\Delta S=\int \Delta C_{vH}/T dT$) between the magnetic and non-magnetic compounds obtained integrating from zero up to the magnetic order temperature ($T_N=30K$ for GdCoIn$_5$ and $T_N=40K$ for GdRhIn$_5$) lies in the range $0.01 J/K\lesssim \vert\Delta S\vert \lesssim 1.9 J/K$ for GdCoIn$_5$, and in the range $0.3 J/K\lesssim \vert\Delta S\vert \lesssim 3.5 J/K$ for GdRhIn$_5$. This difference would lead, in the worst case, to an error of at least $\frac{\vert\Delta S\vert}{R Log(2J+1)} \sim10\%$ ($20\%$) for GdCoIn$_5$ (GdRhIn$_5$) in the magnetic entropy.

\begin{figure}[t]
  \centering
    \includegraphics[width=\textwidth]{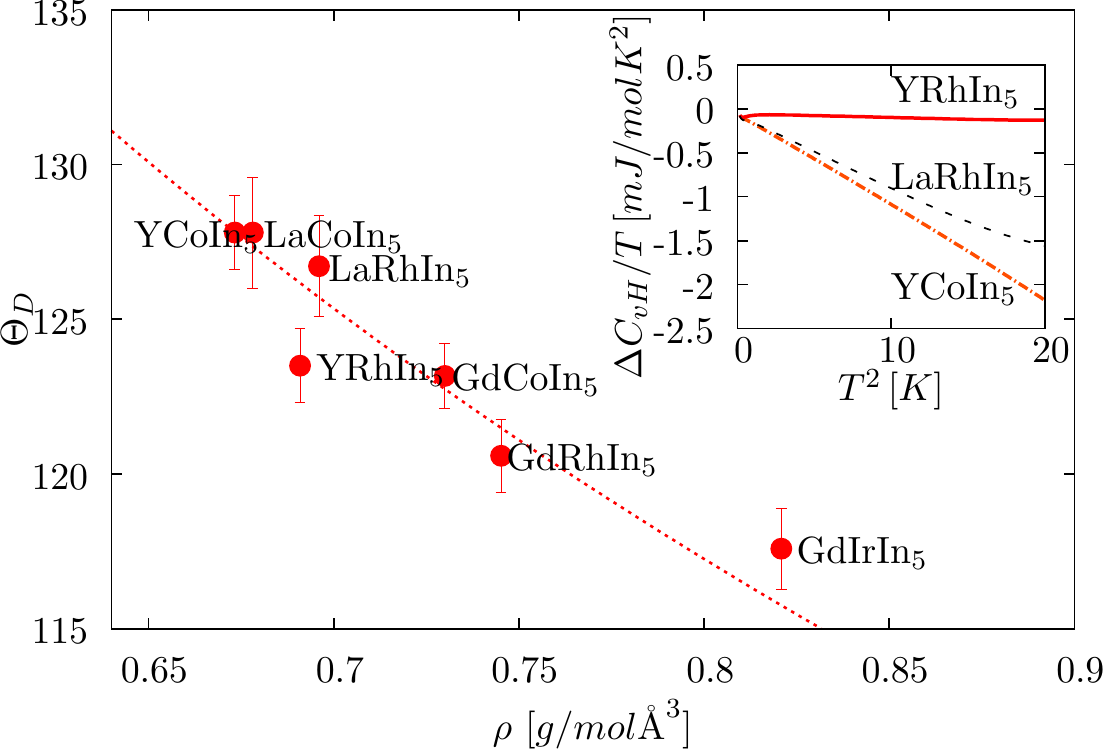}
		\caption{Calculated Debye temperature as a function of the mass density of different 115 compounds. The line is a least squares fit of the data using the function $a_1 \rho^{-1/2}$. The inset presents the difference in specific heat between GdCoIn$_5$ and other compounds (indicated in the figure) at low temperatures. }
  \label{fig:debyeT}
\end{figure}

The low temperature specific heat is dominated by the acoustic modes. For all the compounds studied in this article, the two transverse branches have a lower sound velocity $c_s$ that the longitudinal branch, and provide a much higher contribution to the low temperature specific heat. Within the Debye model, the phonon contribution to the low temperature specific heat is given by
\begin{equation}
    C_D(T)=\beta T^3.
    \label{cdebye}
\end{equation}
where $\beta$ is a function of the Debye temperature $\theta_D$.
We calculated the Debye temperature through a linear fit to $C_{vH}/T$ with a quadratic function $T^2$ (see inset to Fig. \ref{fig:debyeT}). The results are presented in Fig. \ref{fig:debyeT} as a function of the mass density $\rho$ of each compound. Most of the variation in $\theta_D$ between compounds can be explained by the change 
of the mass density: $\theta_D\propto c_s\propto 1/\sqrt{\rho}$.

\subsection{Anharmonic correction} \label{subs:anco}
In our calculations of the phononic contribution to the specific heat, we assumed a temperature independent lattice parameter (constant volume), while our experiments were performed at constant ambient pressure. The constant pressure specific heat $C_p$ is connected to the specific heat at constant volume $C_v$ through the thermodynamical equation (see e.g. Ref. \cite{gaskell2008introduction}):
\begin{equation}
	C_p-C_v = T V_m \alpha_v^2 B_T 
	\label{eq:cpcv1}
\end{equation}
where $B_T$, $V_m$ and $\alpha_v$ are the isothermal bulk modulus, the molar volume and the coefficient of volume thermal-expansion, respectively. We estimated the experimental $C_v$ using the measured $C_p$, $\alpha_v$ and the zero-temperature value $B_T=78GPa$ obtained using DFT. 
We obtain a linear discrepancy at high temperatures $dC_V/dT \sim 40mJ/K^2$ which, as we show below, can be attributed to anharmonic effects (not included in the calculation) that produce a softening of the phononic modes as the temperature is raised.  Following Wallace\cite{Wallace} in the quasiharmonic approximation we assume that the frequency of the phonon modes has a temperature and volume dependence given by $\Omega_k=\Omega_k^0(V)+\Delta_k(T)$, where $\Omega^0_k(V)$ is the harmonic contribution and $\Delta_k(T)$ is an anharmonic contribution. This leads to the following relation
\begin{equation}
	C_v-\hat{C}_{vH} = -T V_m \alpha_v^2 B_T -T\sum_k \left(\frac{\partial \Omega_k}{\partial T}\right)_P \left(\frac{\partial n_k}{\partial T}\right)_V
	\label{eq:Wallace}
\end{equation}
where $\hat{C}_{vH}$ is the harmonic contribution to the specific heat calculated using the shifted phonon frequencies. Assuming the same linear shift $\Omega_k(T)=\Omega_k^0-cT \Omega_k^0$ for all phonon modes, 
\begin{equation}
	C_v + T V_m \alpha_v^2 B_T = \hat{C}_{vH} + c T C_{vH}
	\label{eq:cvcvh}
\end{equation}
replacing $\hat{C}_{vH}$ by $C_{vH}$ to lowest order in $cT$, and combining with Eq.(\ref{eq:cpcv1}) we obtain \cite{Martin1991a}
\begin{equation}
	C_p = C_{vH}(1 + c T)
	\label{eq:cpcvan}
\end{equation}
This linear correction at high temperatures due to anharmonic effects is in accordance with the general anharmonicity theory \cite{LandauSatatMech}.
The anharmonic correction is expected to be negligible at low temperatures. While Eq. (\ref{eq:cpcvan}) is not expected to be valid at low temperatures, we will use it in the full experimental temperature range as its contribution at low temperatures is very small.

\section{Experimental details and sample preparation}\label{sec:exp}
Single crystalline samples of GdCoIn$_5$ and GdRhIn$_5$ were grown by the self flux technique starting from high purity elements as described elsewhere \cite{Betancourth2015}.
Crystal quality and composition were checked by x-ray diffraction (XRD) and energy-dispersive x-ray spectroscopy (EDS), respectively. 
The XRD reflections were successfully indexed with a tetragonal unit cell (HoCaGa$_5$). Lattice parameters for GdCoIn$_5$ are given in Table 1. For GdRhIn$_5$ our values (c= 7.535(5), a=4.706(5)) are slightly larger than those reported in the literature \cite{VanHieu2007}.
Specific heat measurements were performed in both a commercial Quantum Design
PPMS and a silicon nitride (SiN) membrane microcalorimeter \cite{denlinger1994thin} using a standard relaxation technique. Sample typical masses are in the order of a few milligrams.


\section{Theory vs. experiment}\label{sec:comp}
\begin{figure}[t]
    \begin{center}
        \includegraphics[width=\textwidth]{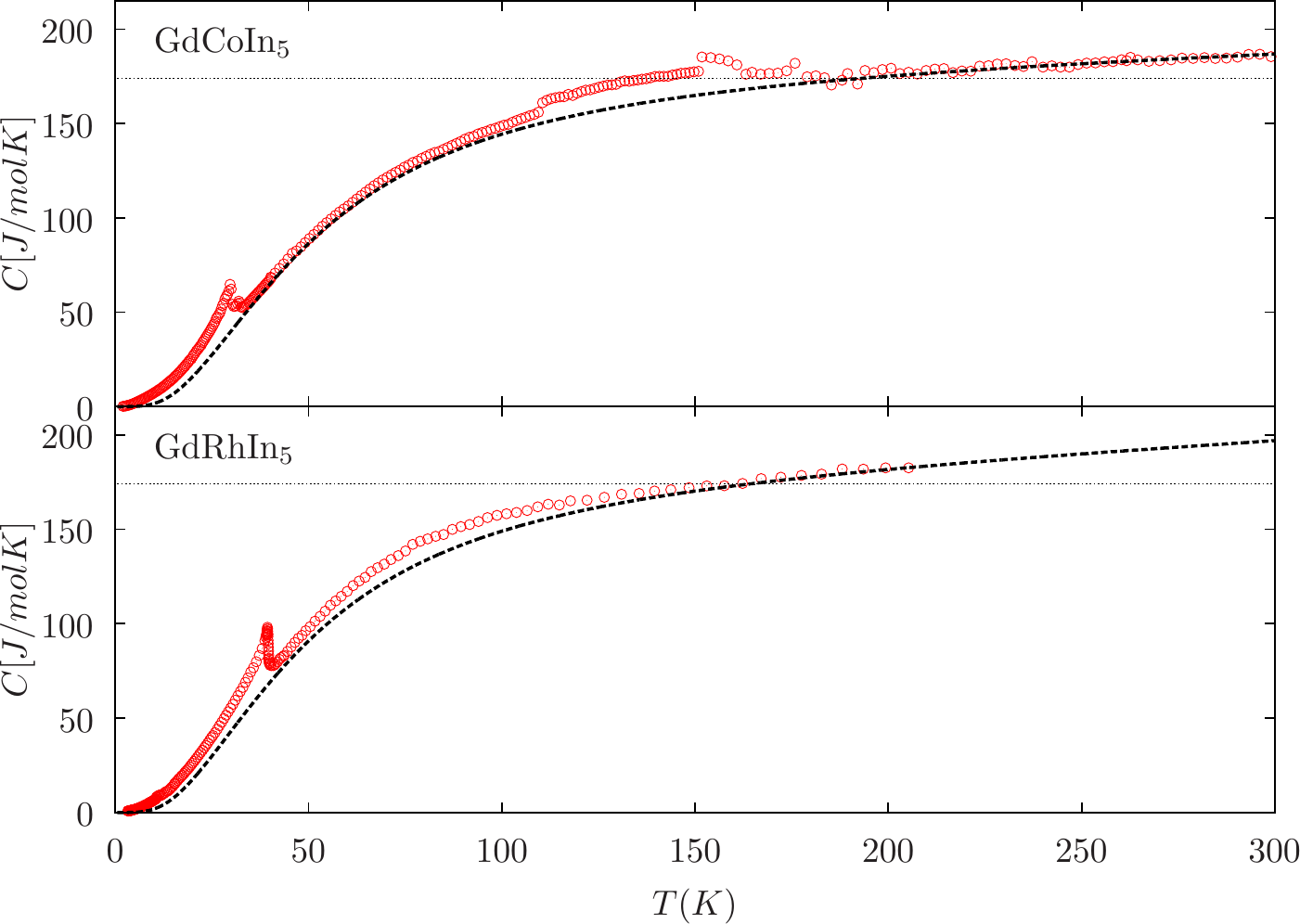}
    \end{center}
    \caption{Top panel: Measured specific heat of GdCoIn$_5$ (circles) and calculated lattice contribution to the specific heat including an anharmonic correction with parameter $c=0.0003$ (see text). The dotted line is the Dulong-Petit value. Bottom panel: Same as top panel for GdRhIn$_5$ and $c=0.00045$.}
    \label{fig:experim}
\end{figure}

Figure \ref{fig:experim} presents the total specific heat measurements for the compounds GdCoIn$_5$ and GdRhIn$_5$. The peak at $T\sim 30K$ for GdCoIn$_5$ and at $T\sim 40K$ for GdRhIn$_5$ is associated with the N\'eel transition. In the high temperature range $T\gtrsim 150K$ the specific heat surpasses the value expected from the Dulong-Petit law (DP) value  expected at high temperatures. The excess from the DP value cannot be explained by magnetic fluctuations~\cite{facio2015co} nor by the electronic contribution as the $\gamma$ value in these materials does not exceed $7mJ/K^2$. We used Eq. (\ref{eq:cpcvan}) to calculate $C_p$ from $C_{vH}$ and include the anharmonicity correction. The value obtained for the anharmonicity coefficient $c$ is $3.0\times 10^{-4} K^{-1}$ for GdCoIn$_5$ and $4.5\times 10^{-4}K^{-1}$ for GdRhIn$_5$. Similar values ($\sim 1\times 10^{-4}K^{-1}$) were reported for the compounds R$_2$RhIn$_8$ (R=Y, La Lu) of the 218 family \cite{vcermak2013specific}. 

\section{Conclusions}\label{sec:concl}
We performed a theoretical and experimental study of the specific heat of 115 materials and analyzed two alternative ways to obtain the magnetic contribution to the specific heat: i) Subtract the specific heat of a non-magnetic analog. ii) Subtract the specific heat calculated using {\it ab initio} methods.
The main difficulty of the experimental subtraction is that the phonon contributions of the analog material and the magnetic compound are different. The substitution of an atomic species to eliminate the magnetism has associated changes in the mass of the atom and in the lattice elastic constants which results in a modified temperature dependence of the specific heat. An incorrect choice of the non-magnetic analog can lead to an error larger than 10\% in the magnetic entropy.
Performing a DFT study of non-magnetic compounds we show that candidates for the subtraction can be found that would produce a very low error in the range of temperatures of interest. The non-magnetic compound to use depends however on the magnetic material and the DFT analysis can be used to guide the choice of a proper
non-magnetic material for the experimental subtraction.
The DFT results for the phonon specific heat can be used directly to perform the subtraction, but we find that there are important anharmonic contributions that need to be taken into account in the high temperature (T> 100K) regime. We find, from an analysis of our experimental results for GdCoIn$_5$ and GdRhIn$_5$, that the anharmonic effects can be taken into account using a single pseudophenomenological parameter.

\section*{Acknowledgements}
This work was partially supported by SeCTyP-UNCuyo 06C347 and CONICET grants PIP00832, PIP00273, and PICT 2012-1069.
\bibliography{phon,115b,devictor}{}
\bibliographystyle{elsarticle-num} 

\end{document}